\documentclass[journal]{IEEEtran}
\usepackage{epsfig,url,multirow,colortbl}
\usepackage[english]{babel}
\usepackage[latin1]{inputenc}
\usepackage{amsmath}
\usepackage{amsfonts}
\usepackage{setspace}
\usepackage{array}
\usepackage{amssymb}
\usepackage{graphicx}
\usepackage{url}
\usepackage{multicol}
\usepackage{stfloats}
\usepackage{subfigure}
\usepackage{booktabs}
\usepackage{changes}
\usepackage{eurosym}
\usepackage{rotating}
\usepackage{nth}
\usepackage{longtable}
\usepackage{color}

\begin{document}

\title{How Far is Facebook from Me? \\Facebook Network Infrastructure Analysis}

\author{\IEEEauthorblockN{
    Reza Farahbakhsh\IEEEauthorrefmark{1},
    Angel Cuevas\IEEEauthorrefmark{1}\IEEEauthorrefmark{2},
    Antonio M. Ortiz\IEEEauthorrefmark{1}\IEEEauthorrefmark{3},
    Xiao Han\IEEEauthorrefmark{1},
    Noel Crespi\IEEEauthorrefmark{1}}\\
  \IEEEauthorblockA{\IEEEauthorrefmark{1}
    Institut Mines-Telecom, Telecom Sud-Paris, France.\\
    E-mail: \{reza.farahbakhsh, han.xiao, noel.crespi\}@it-sudparis.eu} \\
  \IEEEauthorblockA{\IEEEauthorrefmark{2}
    Universidad Carlos III de Madrid, Spain.
    E-mail: {acrumin@it.uc3m.es}}\\
  \IEEEauthorblockA{\IEEEauthorrefmark{3}
    Montimage, Paris, France. E-mail: {antonio.ortiz@montimage.com}}\\
}

\maketitle              

\begin{abstract}
Facebook (FB) is today the most popular social network with more than one billion subscribers worldwide. To provide good quality of service (e.g., low access delay) to their clients, FB relies on Akamai which provides a worldwide content distribution network with a  large number of edge servers that are much closer to FB subscribers. In this paper we aim at depicting a global picture of the current FB network infrastructure deployment taking into account both native FB servers and Akamai nodes. Towards this end, we have performed a measurement based analysis during a period of two weeks using 463 PlanetLab nodes distributed across 41 different countries. Based on the obtained data we compare the average access delay that nodes in different countries experience accessing both native FB servers and Akamai nodes. In addition, we obtain a wide view of the deployment of Akamai nodes serving FB users worldwide. Finally, we analyze the geographical coverage of those nodes, and demonstrate that in most of the cases Akamai nodes located in a particular country not only service local FB subscribers, but also FB users located in nearby countries.
\end{abstract}

\textbf{keywords:} Facebook, Akamai, CDN, Geolocation, Access Delay.


\section{Introduction}

Facebook (FB) is the most popular On-line Social Network (OSN) with more than 1 billion subscribers all over the world.
According to Alexa Ranking\footnote{http://www.alexa.com/topsites}, FB is the 2nd most popular website in the world. A system of that dimension needs to be sustained by a robust and reliable architecture. Toward this end, FB owns and manages a number of centralized datacenters located in the US and Ireland \cite{wongyai2012examining}. However, those datacenters are far from a large number of FB subscribers, who could incur very high delays to reach them.
Access delay is a very sensitive parameter that impacts user experience and may have a big negative effect on online services if it is not bounded.  Some illustrative examples of the actual relevance of delay reported in \cite{bruce_delay} are:  $(i)$ 100ms delay increment implies 1\% sales loss for Amazon, $(ii)$ an extra latency of 400ms reduces Google search volume by 0.74\%, $(iii)$ for Bing 500ms of delay decrements the revenue per user in 1.2\%. These numbers state that the lower the delay the better the quality of experience of the users.

Therefore, to provide an efficient service, a worldwide popular system like FB needs to rely on a distributed infrastructure that provides subscribers a good quality of service (e.g., low access delay). To achieve this goal FB uses Akamai\footnote{http://www.akamai.com}, a Content Distribution Network (CDN) with 170k servers deployed in 102 countries which delivers between 15-30\% of all Web traffic\footnote{Akamai Facts \& Figures, 2014, www.akamai.com/html/about/facts\_figures.html}. 

In this context, an intriguing question that motivates our research is how this complex infrastructure offers FB services\footnote{We refer as FB service to any activity that a regular FB subscriber can perform when she/he is connected to FB like visualizing pictures, watching videos, gaming, chatting, etc.} to FB subscribers, and whether all countries are experiencing the same quality of service in terms of their delay in accessing those services. The goal of this paper is to present a rigorous measurement study that allows us to construct the actual FB infrastructure (including Akamai servers) and see how it is being used to meet subscribers demand. 

To answer the previous question it is essential to determine how the Akamai servers that offer FB services are distributed around the world, and to which Akamai locations FB subscribers are redirected when they access a particular service. Towards this end, we followed a systematic methodology that allows us to identify which Akamai servers are offering what FB services as well as geolocating them.
This methodology is composed of four basic steps: $(i)$ identify the URLs associated with FB services, $(ii)$ execute ping and traceroute commands from edge machines distributed worldwide to extract IP addresses associated with servers attending queries related to the discovered FB services, $(iii)$ geolocate those IPs and determine which ones are associated to native FB servers and which ones belong to Akamai servers, and $(iv)$ determine which source nodes (in which location) are assisted by which Akamai servers.
To apply this methodology we used 463 PlanetLab (PL) \cite{planetlab} nodes distributed across 41 countries all over the world, which sent ping and traceroute probes to 47 different FB URLs (grouped into 16 different service categories) six times a day for two weeks, from May 7th to May 21st, 2013. Overall we collected almost 2M delay samples from PL nodes to FB native servers and Akamai nodes.

Based on the results obtained from our measurements, we present a discussion that mainly covers two aspects: $(i)$ the quality of service (in terms of delay) experienced by subscribers depending on their location,  and $(ii)$ the picture of where Akamai nodes offering access to FB services are located and which geographical areas they cover (i.e., whether an Akamai node located in country $A$ only receives queries from nodes located in that country or if it also serves nodes in other countries, and in such cases whether these are neighbouring countries or not).


The results of our research serve as a solid benchmark to understand the performance offered by CDNs to large demanding clients with hundred of millions of subscribers distributed all over the world.  Therefore, researchers aiming to improve CDN services could use the results presented in the paper to validate their solutions with respect to the performance offered by the largest commercial CDN. In addition, it opens a door to the networking community to analyze what are the main sources of delay in order to propose solutions that minimize end users access delay to services like FB. Finally, the simple yet efficient methodology employed in the paper can be replicated with other online sites and CDNs to perform comparative analysis to our work.

\begin{table}[t!]
  \centering
  \scriptsize
  \caption{Facebook Service Categories, number of URLs for each Service, and the Service Provider (Facebook and/or Akamai).}
    \begin{tabular}{lcl}
    \toprule
    \textbf{Service Category} & \textbf{\#URLs} & \textbf{Service Provider} \\
    \midrule
    \multicolumn{1}{l}{Access website} & 2     & \multicolumn{1}{l}{Facebook \& Akamai} \\
    \multicolumn{1}{l}{Authentication} & 4     & \multicolumn{1}{l}{Facebook \& Akamai} \\
    \multicolumn{1}{l}{Blog site} & 1     & \multicolumn{1}{l}{Facebook} \\
    \multicolumn{1}{l}{Chat} & 2     & \multicolumn{1}{l}{Facebook} \\
    \multicolumn{1}{l}{Developer site} & 1     & \multicolumn{1}{l}{Facebook} \\
    \multicolumn{1}{l}{Error} & 1     & \multicolumn{1}{l}{Facebook} \\
    \multicolumn{1}{l}{Friend finder} & 1     & \multicolumn{1}{l}{Akamai} \\
    \multicolumn{1}{l}{Friend site} & 1     & \multicolumn{1}{l}{Facebook} \\
    \multicolumn{1}{l}{Game applications} & 3     & \multicolumn{1}{l}{Facebook} \\
    \multicolumn{1}{l}{Group site} & 1     & \multicolumn{1}{l}{Facebook} \\
    \multicolumn{1}{l}{Multi-services} & 4     & \multicolumn{1}{l}{Facebook \& Akamai} \\
    \multicolumn{1}{l}{New Feed} & 4     & \multicolumn{1}{l}{Facebook} \\
    \multicolumn{1}{l}{Photo upload} & 1     & \multicolumn{1}{l}{Facebook} \\
    \multicolumn{1}{l}{Photo view} & 19    & \multicolumn{1}{l}{Facebook \& Akamai} \\
    \multicolumn{1}{l}{Post site} & 1     & \multicolumn{1}{l}{Facebook} \\
    \multicolumn{1}{l}{Video view} & 1     & \multicolumn{1}{l}{Akamai} \\
    \bottomrule
    \end{tabular}%
  \label{tab:services}%
\end{table}

\section{Methodology}
The goal of this article is twofold:
$(i)$ to analyze the user experience in accessing FB services from different countries in terms of latency, and
$(ii)$ to describe a geographical picture for the location of those servers (with an special focus on Akamai nodes) offering FB services, and, linked to that location, whether they only cover a local region or if they also serve users located in  different countries.
Towards this end, we have employed a simple yet meaningful methodology that could be replicated to evaluate the performance in terms of access delay that a CDN offers to a particular website. Next, we define in detail the steps followed in our methodology:

\begin{figure}[t]
	\centering
	\includegraphics[width=0.48\textwidth, height=0.26\textwidth]{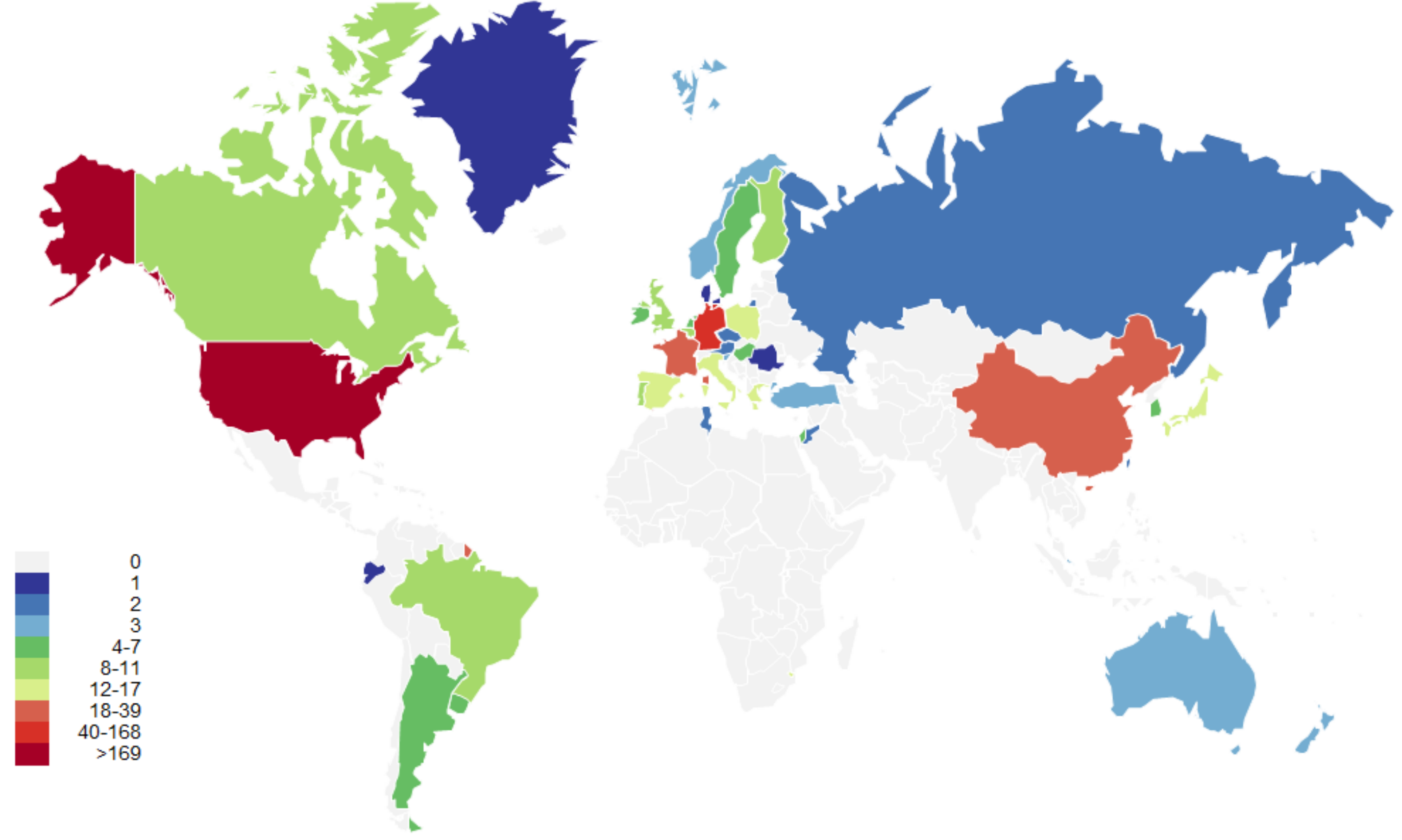}
	\caption{Presence and distribution of the 463 PlanetLab nodes (PL\_node) per country.}
	\label{fig:country_map}
\end{figure}

\begin{table}[t]
  \centering
  \scriptsize
  \caption{Distribution of the 463 PlanetLab nodes (PL\_node) per country.}
  \scalebox{.99}{%
    \begin{tabular}{llr|llr}
\toprule
    \textbf{Country} & \textbf{Acr.}  & \textbf{\#PL\_node} & \textbf{Country} & \textbf{Acr.}  & \textbf{\#PL\_node} \\
    \midrule
    United States & US    & \multicolumn{1}{c|}{169} & Argentina & AR    & \multicolumn{1}{c}{4} \\
    Germany & DE    & \multicolumn{1}{c|}{40} & Hungary & HU    & \multicolumn{1}{c}{4} \\
    China & CN    & \multicolumn{1}{c|}{19} & Korea, Rep. & KR    & \multicolumn{1}{c}{4} \\
    France & FR    & \multicolumn{1}{c|}{18} & Netherlands & NL    & \multicolumn{1}{c}{4} \\
    Italy & IT    & \multicolumn{1}{c|}{16} & Australia & AT    & \multicolumn{1}{c}{3} \\
    Poland & PL    & \multicolumn{1}{c|}{16} & New Zealand & NZ    & \multicolumn{1}{c}{3} \\
    Spain & ES    & \multicolumn{1}{c|}{16} & Norway & NO    & \multicolumn{1}{c}{3} \\
    Greece & GR    & \multicolumn{1}{c|}{12} & Singapore & SG    & \multicolumn{1}{c}{3} \\
    Japan & JP    & \multicolumn{1}{c|}{12} & Slovenia & SI    & \multicolumn{1}{c}{3} \\
    Switzerland & SZ    & \multicolumn{1}{c|}{12} & Turkey & TR    & \multicolumn{1}{c}{3} \\
    Canada & CA    & \multicolumn{1}{c|}{11} & Austria & AT    & \multicolumn{1}{c}{2} \\
    United Kingdom & UK    & \multicolumn{1}{c|}{11} & Czech Rep. & CZ    & \multicolumn{1}{c}{2} \\
    Belgium & BE    & \multicolumn{1}{c|}{9} & Jordan & JO    & \multicolumn{1}{c}{2} \\
    Brazil & BR    & \multicolumn{1}{c|}{8} & Puerto Rico & PR    & \multicolumn{1}{c}{2} \\
    Finland & FI    & \multicolumn{1}{c|}{8} & Russia & RU    & \multicolumn{1}{c}{2} \\
    Portugal & PT    & \multicolumn{1}{c|}{8} & Taiwan & TW    & \multicolumn{1}{c}{2} \\
    Israel & IL    & \multicolumn{1}{c|}{6} & Tunisia & TN    & \multicolumn{1}{c}{2} \\
    Sweden & SE    & \multicolumn{1}{c|}{6} & Denmark & DK    & \multicolumn{1}{c}{1} \\
    Hong Kong & HK    & \multicolumn{1}{c|}{5} & Ecuador & EC    & \multicolumn{1}{c}{1} \\
    Ireland & IE    & \multicolumn{1}{c|}{5} & Romania & RO    & \multicolumn{1}{c}{1} \\
    Uruguay & UY    & \multicolumn{1}{c|}{5} &       &       &  \\
\hline
    \end{tabular}%
    }
  \label{tab:num_node_country}%
\end{table}

\noindent \textbf{Step 1 - Identify URLs associated to the service offered by the website (i.e., FB):}
We asked several Facebook subscribers to perform a number of activities in Facebook such as: login in the site, access their profile, access photos and videos, access friends content, etc.
In parallel, we used a network protocol analyzer tool \cite{wireshark} that collected all the traffic associated to each of the described actions. After a simple filtering of the network traces we could map each Facebook action to one (or more) URLs that could refer either to a FB native server (e.g. profile.facebook.com) or an Akamai server (e.g. photos-a.ak.fbcdn.net). We identified 47 URLs that correspond to 16 different FB services.
To be sure that the URLs were not location-dependant we repeated this exercise in several machines at different geographical locations leading to the same results. Table \ref{tab:services} shows the 16 identified FB service categories included in this study as well as the information on which service provider, Akamai and/or FB, is in charge of replying to the queries for these services.

\noindent \textbf{Step 2 - Script to measure access delay and network path to the URLs:} We implemented a simple script, following a standard discovery method \cite{Internet_Topology}, that executes ping and traceroute operations from the machine where it is executed to all the identified 47 URLs. The ping measures the latency from the source node to the queried server, which serve us to evaluate the performance in terms of access delay. The traceroute reports the intermediate hops between the source node and the server and the delay to each hop (in case the intermediate router accepts ICMP traffic). The traceroute results may serve to dig into the particular reasons of why a particular source node-server path is incurring in unexpected delays and try to identify the elements in the paths leading to that situation. However, that individualized analysis goes beyond the scope of this paper and would require a paper itself.

\noindent \textbf{Step 3 - Create a distributed infrastructure to obtain comprehensive results from different geographic locations:} The goal of this research required to measure access delay to the servers serving the 47 URLs from a large number of source machines distributed all over the world. For this purpose we relied on PlanetLab (PL) \cite{planetlab}. In particular, we distributed the script described in Step 2 across 463 PL nodes located in 41 different countries (see Figure \ref{fig:country_map}) as shown in Table \ref{tab:num_node_country}. In addition, in order to have a large enough and robust dataset that avoids eventual network effects that could corrupt the average delay results, we ran the script 6 times a day (every 4 hours at the same time across all machines) in each PL node during a period of two weeks from May 7th to May 21st, 2013. Our dataset contains more than 2M ping and traceroute probes.

\noindent \textbf{Step 4 - Source Nodes, FB servers and Akamai Servers geolocation}: Until this step we have a large dataset in which each ping probe is associated to a source IP address (i.e., PlanetLab node), destination IP address (i.e., FB or Akamai server) and delay. However, in order to perform the study described in the introduction we have to geolocate each IP address so that for each ping entry in our dataset we also know location of source node and location of destination node. To geolocate each source node, FB server and Akamai server we used the Maxmind database\footnote{http://www.maxmind.com/} to bind each IP address to its respective location. The location included country and city (if available).

We would like to note that the final dataset employed in our research is publicly available for the research community\footnote{http://www.it.uc3m.es/acrumin/papers/FB\_Arch\_project.rar}.


\section{End Users' Access Delay to Facebook Services}
\label{sec:access_delay}

In this section we aim to understand the performance level experienced by end users in terms of the latency in accessing FB services located either in native FB or Akamai servers.
Table \ref{tab:delay_hops_facebook} shows the detail of the average access delay (and its standard deviation) per country to access FB services in servers located in FB facilities, and Table \ref{tab:delay_hops_akamai} shows the same parameters in relation to Akamai servers.
In addition Figure \ref{fig:Figure_delay} show the average access delay to access FB services in servers located at FB facilities (Figure \ref{fig:delay_hops_FB}) and at Akamai facilities (Figure \ref{fig:delay_hops_AK}). 
Overall, FB users need 113ms in average to access native FB servers, but only 43ms to reach Akamai nodes providing FB access. This means that accessing FB services in Akamai nodes reduces 2.5$\times$ the delay. Next, we provide a detailed analysis of the access delay performance per country.

\subsection{Access delay to native Facebook servers}
\begin{figure}[t]
	\centering
\subfigure[{Facebook}]
{\includegraphics[width=0.48\textwidth, height=0.26\textwidth]{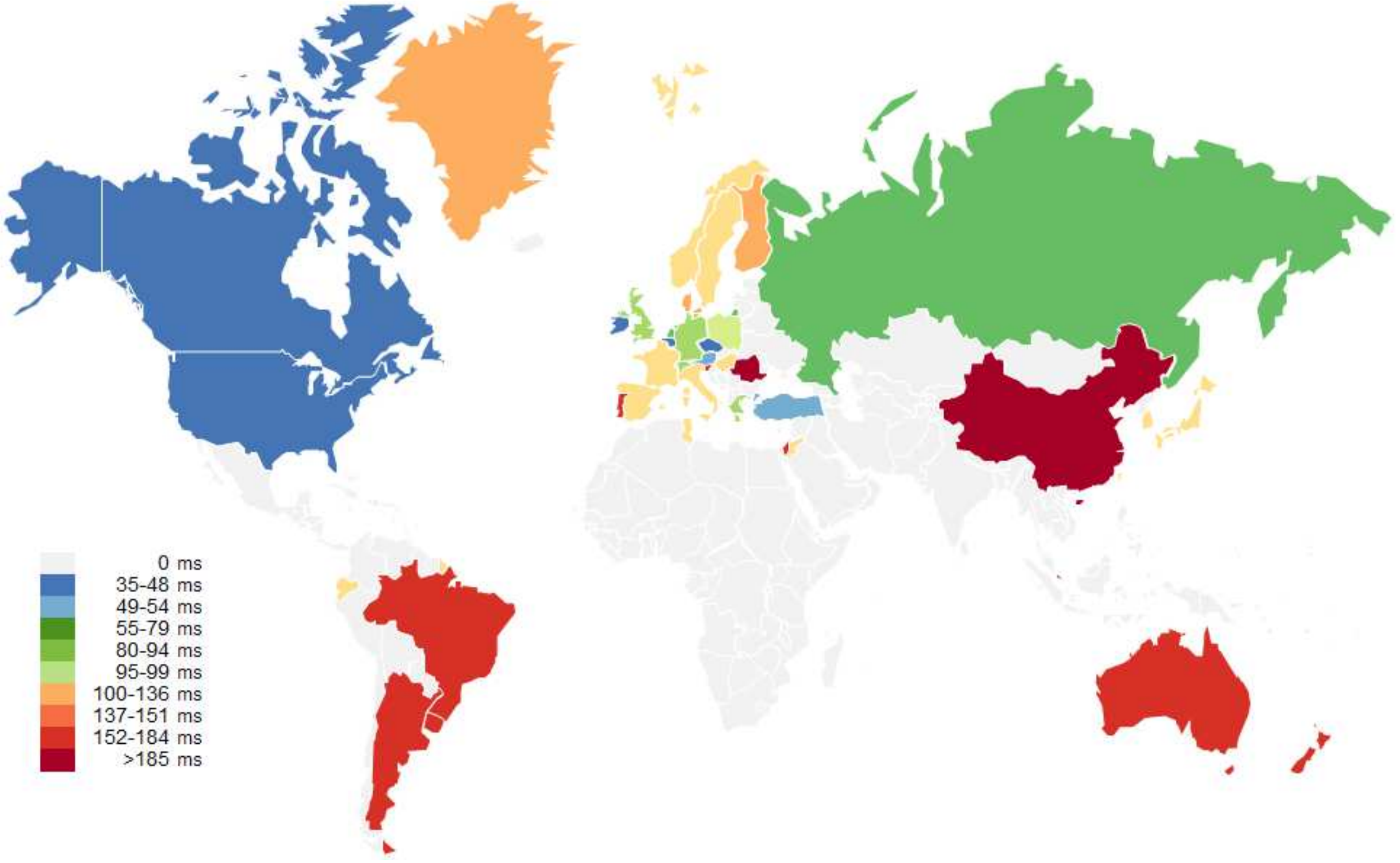}
\label{fig:delay_hops_FB}}
\subfigure[{Akamai}]
{\includegraphics[width=0.48\textwidth, height=0.26\textwidth]{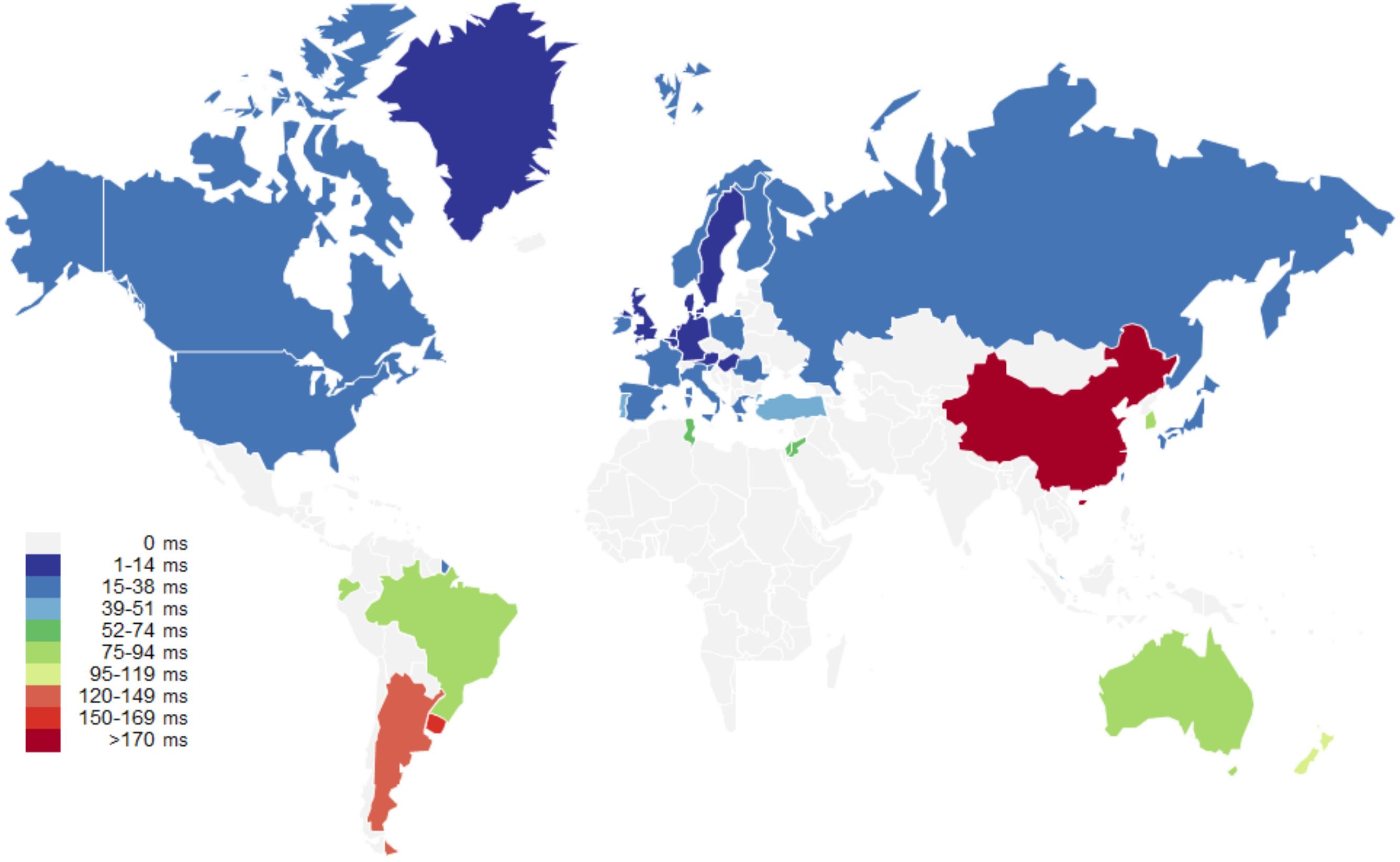}	
\label{fig:delay_hops_AK}}
\caption{Average delay (Miliseconds) to access FB services from different countries for services
    (a) located in FB serves
    (b) located in Akamai servers.}
    \label{fig:Figure_delay}
\end{figure}

Based on the results of Table \ref{tab:delay_hops_FB_AK}, we have defined four groups in terms of their access delay to FB servers which are illustrated in different color range in Figure \ref{fig:Figure_delay} as well.

$(1)$ The first group refers to all those countries with an access delay longer than 150ms Red group in Figure \ref{fig:delay_hops_FB}). This group is formed by countries that are quite far from US (e.g., Australia, New Zealand), South-American countries, and three countries we did not expect to find in this group such as Portugal, Slovenia and Israel since their surrounding neighbours show a  considerably lower delay.

$(2)$ The second group is formed of those countries whose delay ranges between 100ms and 150ms Orange group). This group includes Northern European countries, Asian countries with a deep penetration of high speed access connections (e.g., Japan, South Korea, Hong Kong), countries from central America and Mediterranean countries including some important European parties like France, Italy and Spain.

$(3)$ The third group includes those countries with a delay greater than 50ms but less than 100ms (Green group). This group is mainly formed by countries located in Central Europe plus Greece, Turkey and the UK.

$(4)$ The last group contains those countries with access delay is under than 50ms (Blue group). This includes the two countries hosting native FB servers, the US and Ireland \cite{wongyai2012examining}, and Canada due to its proximity and good connectivity with the US. Surprisingly, this group also includes Belgium and the Czech Republic which intuitively would have fitted better in the third group.

\begin{table}[t!]
\centering
\scriptsize
\caption{Average delay (Miliseconds) $\pm$ Standard Deviation to access FB services from different countries for services located in FB serves \ref{tab:delay_hops_facebook} and services located in Akamai servers \ref{tab:delay_hops_akamai}.}
\label{tab:delay_hops_FB_AK}
\label{fnru435n}
\scalebox{.9}{

\subtable[\textbf{Facebook}]{
\begin{tabular}{lc}
    \toprule
     \textbf{Country}     &  \textbf{Avg.Delay(ms) $\pm$ STD}\\
    \midrule
    \textbf{(1)}&\\
     Singapore       & 193.66 $\pm$ 59.41 \\
     Romania             & 190.07 $\pm$ 50.55 \\
     China               & 187.14 $\pm$ 227.29  \\
     Uruguay             & 179.96 $\pm$ 65.08  \\
     Portugal            & 177.91 $\pm$ 69.02 \\
     Slovenia            & 169.86 $\pm$ 48.50  \\
     Brazil              & 169.78 $\pm$ 60.93 \\
     Israel              & 167.14 $\pm$ 90.85  \\
     Australia           & 164.11 $\pm$ 43.22  \\
     Argentina           & 155.38 $\pm$ 67.49  \\
     New Zealand         & 152.02 $\pm$ 38.00  \\
     \midrule
     \textbf{(2)}&\\
     Denmark         & 140.93 $\pm$ 38.52 \\
     Finland             & 137.12 $\pm$ 61.17  \\
     France              & 133.12 $\pm$ 61.04  \\
     Korea, Rep.  & 128.84 $\pm$ 76.56 \\
     Japan               & 126.96 $\pm$ 64.96  \\
     Sweden              & 114.28 $\pm$ 56.11  \\
     Jordan              & 109.95 $\pm$ 61.85  \\
     Puerto Rico         & 108.42 $\pm$ 36.14  \\
     Ecuador             & 106.69 $\pm$ 36.66  \\
     Tunisia             & 104.47 $\pm$ 50.99 \\
     Norway              & 104.01 $\pm$ 62.21  \\
     Italy               & 102.57 $\pm$ 75.37  \\
     Taiwan              & 101.71 $\pm$ 85.34  \\
     Spain               & 100.94 $\pm$ 73.13  \\
     Hong Kong           & 100.58 $\pm$ 84.43  \\
     Hungary             & 100.05 $\pm$ 76.77  \\
     \midrule
     \textbf{(3)}&\\
     Poland          & 99.69 $\pm$ 58.80  \\
     Greece              & 92.70 $\pm$ 69.36  \\
     UK      & 90.46 $\pm$ 50.67 \\
     Switzerland         & 88.40 $\pm$ 66.13  \\
     Germany             & 84.47 $\pm$ 61.80  \\
     Russia  & 77.49 $\pm$ 52.54  \\
     Netherlands         & 59.52 $\pm$ 54.77 \\
     Austria             & 53.75 $\pm$ 50.77  \\
     Turkey              & 51.37 $\pm$ 64.37  \\
     \midrule
     \textbf{(4)}&\\
     Czech Rep.      & 48.36 $\pm$ 51.95  \\
     Ireland             & 45.88 $\pm$ 50.55  \\
     Belgium             & 42.70 $\pm$ 56.02  \\
     Canada              & 38.51 $\pm$ 46.15  \\
     US       & 36.81 $\pm$ 34.72  \\
    \bottomrule
\end{tabular}
\label{tab:delay_hops_facebook}
}
\subtable[\textbf{Akamai}]{
\begin{tabular}{lc}
    \toprule
    \textbf{Country}     &  \textbf{Avg.Delay(ms) $\pm$ STD}  \\
    \midrule
    \textbf{(1)}&\\
	China & 174.59 $\pm$ 213.30 \\
	Uruguay & 157.40 $\pm$ 78.98\\
	Argentina & 124.98 $\pm$ 79.67  \\
   	 \midrule
   	 \textbf{(2)}&\\
	New Zealand & 95.98 $\pm$ 83.41 \\
	Korea, Rep. & 90.14 $\pm$ 90.75  \\
	Australia & 87.03 $\pm$ 89.32  \\
	Ecuador & 79.62 $\pm$ 55.81  \\	
	Brazil & 78.22 $\pm$ 68.44  \\
	Hong Kong & 71.41 $\pm$ 80.92  \\
	Jordan & 68.72 $\pm$ 38.89  \\
	Tunisia & 63.05 $\pm$ 27.39  \\
	Israel & 54.64 $\pm$ 78.04  \\
    \midrule
    \textbf{(3)}&\\
	Portugal & 49.43 $\pm$ 16.24  \\
	Singapore & 45.32 $\pm$ 73.15  \\
	Puerto Rico & 41.86 $\pm$ 41.65  \\
	Turkey & 39.14 $\pm$ 45.65\\
	Taiwan & 35.74 $\pm$ 57.75 \\
	Greece & 33.78 $\pm$ 24.88 \\
	Japan & 30.42 $\pm$ 42.95 \\
	Spain & 27.25 $\pm$ 19.00 \\
	Russia  & 26.38 $\pm$ 20.41 \\
	Romania & 26.36 $\pm$ 17.35 \\
	Ireland & 24.35 $\pm$ 41.62 \\
	France  & 24.34 $\pm$ 46.36 \\
	Norway & 23.33 $\pm$ 15.62 \\
	Poland & 23.15 $\pm$ 10.31 \\
	Canada & 22.59 $\pm$ 38.57 \\
	Finland & 22.54 $\pm$ 17.42 \\
	Slovenia  & 18.70 $\pm$ 17.11 \\
	US & 15.90 $\pm$ 25.02 \\
	Italy & 15.06 $\pm$ 12.83 \\
	Germany & 10.94 $\pm$ 8.58 \\
	UK & 10.80 $\pm$ 11.74 \\
	Belgium & 10.68 $\pm$ 27.21 \\
	Sweden & 10.20 $\pm$ 10.98 \\
	Hungary & 8.80  $\pm$ 7.36 \\
	Switzerland & 8.56  $\pm$ 12.02  \\
	Netherlands & 7.77  $\pm$ 13.00 \\
	Denmark & 7.09  $\pm$ 6.02  \\
	Austria & 6.84  $\pm$ 5.76  \\
	Czech Rep.& 3.22 $\pm$ 3.64  \\
    \bottomrule
\end{tabular}
\label{tab:delay_hops_akamai}
}}
\end{table}

\subsection{Access delay to Akamai servers}
In the case of Akamai nodes we just define three groups for our discussion.

$(1)$ The first group is formed by three countries that experience an average delay longer than 100ms (Red group in Figure \ref{fig:delay_hops_AK}). These countries are China, Argentina and Uruguay. This happens because an important portion of the FB queries from these countries are redirected to remote Akamai nodes, which could be located for instance in the US.

$(2)$ We formed a second group consisting of countries with an average access delay ranging between 50ms and 100ms (Green group). This include far eastern countries like Australia, New Zealand, South Korea and Hong Kong, two countries in South-America, Brazil and Ecuador, and three countries from North-Africa and the Middle East: Jordan, Tunisia and Israel. As we will see in Section \ref{sec:akamai}, the first six countries count with their own Akamai nodes but a relevant portion of their demand is attended from foreign Akamai servers. In addition, Jordan and Tunisia do not host any Akamai node, but are served by Akamai nodes located in Europe, which is relatively close.
It is surprising that Australia (as a developed country) experience a quite bad performance to access FB services through Akamai nodes. To have a better insight, we leveraged the FB ads planner\footnote{ https://www.facebook.com/ads} to retrieve the potential reach for ads in each country. We have found that Australia has a potential reach of 13M FB users, while some of the countries in the 3rd group, like Greece and Slovenia, which present 50ms and 70ms less average access delay, have a potential reach of 4.4M and 0.7M, respectively.
Another surprising case in this group is Brazil, a huge country with more than 200M population and potential reach of 86M audience from FB ads, while shows an average Akamai access delay around 78ms.

$(3)$ Finally, we define a group including the countries with an access delay below 50ms (Blue group). This group mainly includes developed countries from Europe, Asia (i.e., Japan and Singapore) and North America (US and Canada). This is a good estimation of a short-list of important countries for FB, where FB is interested on offering a better quality of service through Akamai nodes.

Furthermore, it is interesting to note that Akamai offers the best delay performance (i.e., below 10ms) to small countries roughly located in Central Europe (Hungary, Switzerland, Netherlands, Denmark, Austria and Czech Republic). This happens because these are very small countries (in size) that experience a very small delay 
due to the short distance 
to a large number of Akamai nodes located in Central Europe. 

\section{Akamai nodes distribution to provide access to Facebook services}
\label{sec:akamai}
This section provides a global picture of the deployment of Akamai nodes to serve FB services worldwide.

\subsection{Local vs. External Access}

\begin{figure}[t]
	\centering
	\includegraphics[width=0.48\textwidth, height=0.51\textwidth]{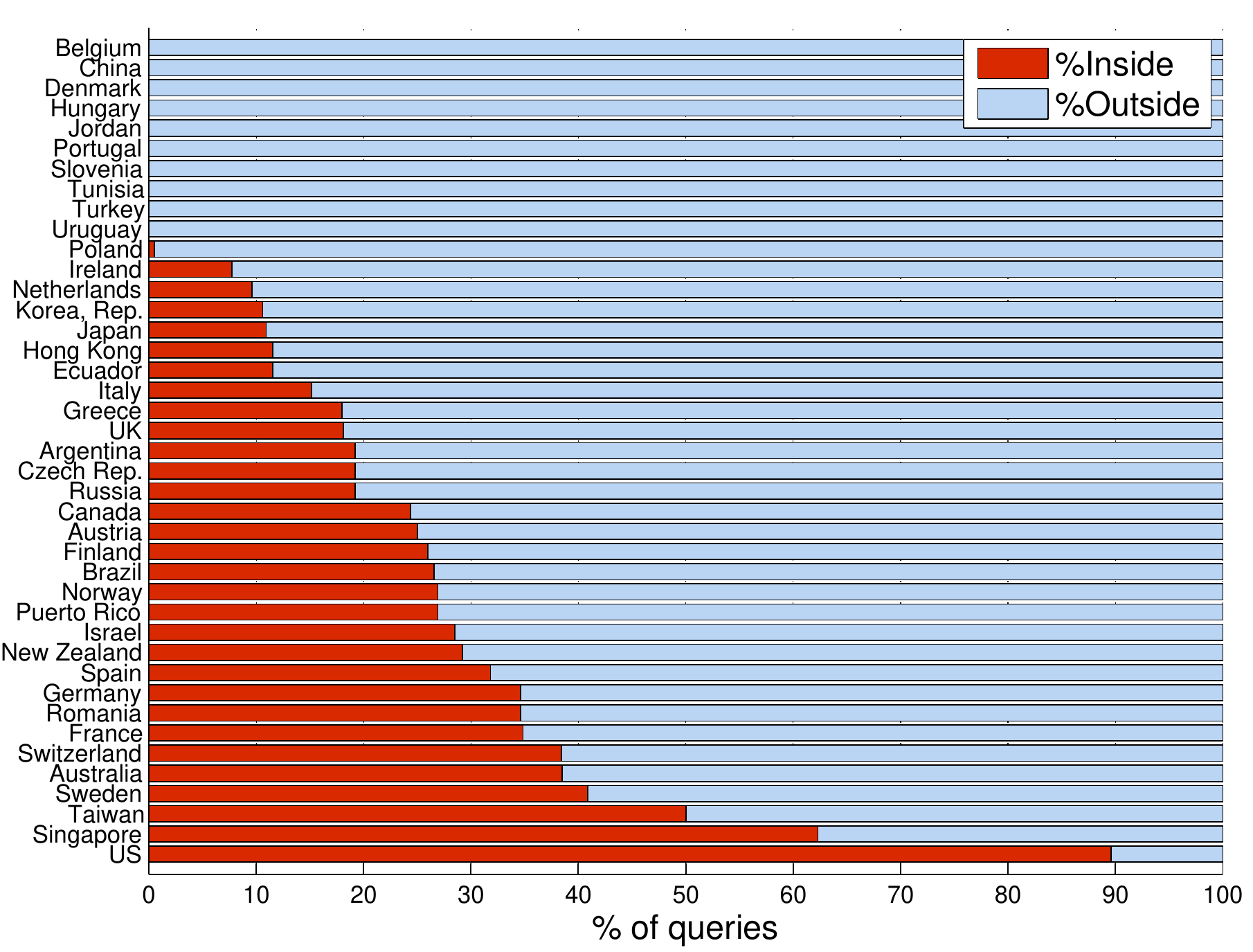}
\vspace{-4mm}
	\caption{Portion of FB queries from each country served by local (\%Inside) and foreign (\%Outside) Akamai nodes.}
	\label{fig:traffic_inside_outside}
\end{figure}

Figure \ref{fig:traffic_inside_outside} shows which portion of the queries for each country (i.e., ICMP echo) is managed by Akamai servers hosted in the same country than the source node(s) and which portion is served by Akamai nodes in foreign countries.
There are only two countries showing a higher portion of local access to Akamai servers compared to external access which are US and Singapore with 90\% and 62\% of the queries going to local Akamai servers. The case of Singapore might be unexpected, but as we will show later, Singapore has a high number of Akamai nodes (i.e., IPs). 
Close to Singapore performance, we find the case of Taiwan in which half of the queries are dealt with local servers and half by foreign servers.

We found that there are a limited number of countries that use local Akamai nodes to serve between 30\%-40\% of their queries. These are:
$(i)$ the largest European countries by size (i.e., Germany, France and Spain) all of which have a large number of Akamai servers.
$(ii)$ Australia, another large country with high number of FB subscribers, that are located far from native FB servers and thus FB is motivated to use Akamai CDN to offer a good performance to Australian subscribers, and
$(iii)$ Three European countries, Switzerland, Sweden and Romania, each particularly distributed geographically in the center, north and east of Europe respectively. 
The Akamai infrastructures in Switzerland and Sweden bring them to have access delays to Akamai nodes in the order of 10ms. Finally, Romania has just six Akamai servers that service 35\% of the queries generated in Romanian nodes. 

Next, we found a large number of countries keeping between 7\% and 30\% of FB queries were responded locally, while most of them were serviced by foreign Akamai servers. Each of these countries have more or less Akamai nodes that allow keeping part of the queries locally, but their delay is mostly affected by how far those Akamai nodes are located from the major part of their queries.

Finally, there were 10 countries for which we could not identify any local Akamai server.
Among these are five European countries (Belgium, Denmark, Hungary, Portugal and Slovenia), each with a population under 10M and close to countries with a significant deployment of Akamai nodes running FB services.
The fact that these countries are experiencing a very good service by accessing Akamai nodes in nearby countries explains the low presence of Akamai servers.

This group of countries without Akamai servers also includes Turkey, 
which we found similar delay to some European countries like Greece or Portugal because the three PL nodes used for our experiments are located in the western part of Turkey (i.e., Istanbul and Izmir).
Next, we discuss the case of Uruguay, a small South-American country surrounded by Argentina and Brazil that already contains some Akamai servers. Interestingly, the results in Table \ref{tab:ips_country_akamai} show that the five PL nodes placed in Uruguay access Akamai servers located in Brazil as well as servers in Mexico and US that are far away, but never go to Argentina.
Two small countries, Tunisia and Jordan, both of them are served by Akamai nodes located (mainly) in Europe. Finally, we find China which is currently blocking FB, and thus it does not make sense to deploy Akamai nodes to serve FB subscribers and they are served by Akamai nodes all over the world.

\subsection{Country coverage by Akamai Servers}
Table \ref{tab:ips_country_akamai} shows for each country which is hosting Akamai nodes, the overall number of IPs linked to Akamai nodes located in that country (column \#IP), and the list of countries hosting nodes that access those IPs\footnote{For simplicity during the discussion we will use the number of IPs as the number of servers/nodes, even though we are aware that it is feasible that the same physical server could hold more than one IP (e.g., multiple network cards, virtualization, etc.)} (column Served To(\#IP)). For each source-querying country we represent the overall number of IPs (between brackets) accessed in the destination country hosting Akamai nodes. 

We found 35 countries that host Akamai nodes to provide FB access to the 41 countries represented by PL nodes. Among them, at the top of Table \ref{tab:ips_country_akamai}, we find 13 countries where Akamai nodes only serve local users. In the middle of the table we list four countries: Azerbaijan, Malaysia, Mexico and Panama, whose Akamai nodes only serve foreign countries. In fact, this behaviour responds to the fact that we did not have any PL node located in those countries. Otherwise, we would very likely have observed that these Akamai nodes also serve local users.
Finally, at the bottom of the table, we find a major part of the countries (18 in total) whose Akamai nodes process queries from both local and foreign PL nodes. Next, we discuss the most interesting aspects for this group.

\begin{table}[t!]
  \centering
  \scriptsize
  \caption{ First column shows the list of countries hosting Akamai nodes offering access to FB services.
  Second column shows the number of identified Akamai-related IPs in each country.
  The third column shows the list of countries including nodes querying Akamai IPs in the country referred in the 1st column.
  The number between brackets reflects the number of IPs accessed in the reference (1st column) country.}
  \begin{center}
  \scalebox{.97}{%

    \begin{tabular}{|l|c|p{5.5cm}|}
\hline
    \multicolumn{1}{|c|}{\textbf{Country}} & \multicolumn{1}{c|}{\textbf{\#IP}} & \multicolumn{1}{l|}{\textbf{Served To(\#IP)}} \\
\hline
    Argentina & 9    & Argentina(9) \\ \hline
    Canada & 28   & Canada(22) \\ \hline
    Ecuador & 3    & Ecuador(3) \\ \hline
    Greece & 7    & Greece(7) \\ \hline
    HongKong & 6    & HongKong(6) \\ \hline
    Israel & 18   & Israel(18) \\ \hline
    Korea & 7    & Korea(7) \\ \hline
    Poland & 2    & Poland(2) \\ \hline
    Puerto Rico & 8    & Puerto Rico(8) \\ \hline
    Romania & 6    & Romania(6) \\ \hline
    Russia & 7    & Russia(7) \\ \hline
    Spain & 35   & Spain(35) \\ \hline
    Taiwan & 9    & Taiwan(9) \\
\specialrule{.2em}{.05em}{.05em}
    Azerbaijan & 1    & China(1) \\ \hline
    Malaysia & 21   & HongKong(3), NewZealand(16), Singapore(2) \\ \hline
    Mexico & 4    & Uruguay(4) \\ \hline
    Panama & 4    & Canada(4) \\
\specialrule{.2em}{.05em}{.05em}
    Australia & 15   & Australia(10), Japan(4), Taiwan(1) \\ \hline
    Austria & 49   & Austria(9), Greece(24), Hungary(26), Israel(2), Poland(37), Slovenia(27) \\ \hline
    Brazil & 26   & Brazil(22), Uruguay(16) \\ \hline
    Czech & 11   & Czech(6), Poland(7), Russia(4) \\ \hline
    Finland & 24   & Finland(19), Norway(4), Russia(3), Sweden(12) \\ \hline
    France & 176  & Belgium(20), Finland(4), France(60), Germany(4), Greece(1), Hungary(1), Ireland(7), Israel(3), Jordan(20), Poland(10), Singapore(3), Spain(41), Switzerland(5), Tunisia(4), Turkey(1), UnitedKingdom(48) \\ \hline
    Germany & 473  & Australia(3), Austria(11), Belgium(30), China(6), Czech(13), Denmark(4), Finland(19), France(25), Germany(184), Greece(43), Hungary(11), Ireland(9), Israel(20), Italy(20), Jordan(5), Netherlands(12), Norway(5), Poland(31), Portugal(24), Romania(7), Russia(12), Slovenia(2), Spain(47), Sweden(11), Switzerland(86), Tunisia(9), Turkey(16), UnitedKingdom(14), UnitedStates(4) \\ \hline
    Ireland & 6    & China(1), Ireland(5) \\ \hline
    Italy & 49   & China(1), Greece(4), Hungary(2), Israel(1), Italy(20), Jordan(14), Switzerland(1), Tunisia(6), Turkey(1), UnitedStates(2) \\ \hline
    Japan & 36   & China(17), HongKong(4), Japan(16), Korea(5) \\ \hline
    Netherlands & 39   & Belgium(3), China(1), France(4), Ireland(14), Netherlands(6), Tunisia(6), UnitedKingdom(1), UnitedStates(3) \\ \hline
    NewZealand & 11   & China(1), NewZealand(10) \\ \hline
    Norway & 8    & Finland(2), Norway(5), Sweden(2) \\ \hline
    Singapore & 110  & Argentina(3), Brazil(26), China(2), Ecuador(4), HongKong(13), Japan(1), Korea(3), NewZealand(1), Puerto Rico(15), Singapore(26), Taiwan(1), UnitedStates(30), Uruguay(12) \\ \hline
    Sweden & 77   & Denmark(1), Finland(31), Ireland(7), Norway(6), Poland(7), Russia(10), Sweden(24), UnitedKingdom(19) \\ \hline
    Switzerland & 49   & Australia(5), Poland(9), Sweden(8), Switzerland(33) \\ \hline
    UnitedKingdom & 246  & Belgium(21), Denmark(4), France(19), Germany(22), Greece(34), Hungary(7), Ireland(17), Israel(34), Italy(2), Netherlands(23), Norway(13), Poland(29), Portugal(21), Romania(1), Spain(21), Sweden(5), Switzerland(11), Tunisia(9), Turkey(15), UnitedKingdom(45), UnitedStates(4) \\ \hline

  \multicolumn{1}{|l|}{UnitedStates} & 2505 & Argentina(67), Australia(27), Austria(19), Belgium(39), Brazil(48), Canada(148), China(177), Czech(17), Denmark(13), Ecuador(17), Finland(16), France(69), Germany(117), Greece(32), HongKong(67), Hungary(26), Ireland(16), Israel(11), Japan(96), Jordan(1), Korea(66), Netherlands(18), NewZealand(21), Norway(17), Poland(47), Portugal(79), Puerto Rico(17), Romania(7), Singapore(21), Slovenia(14), Spain(24), Sweden(22), Switzerland(9), Taiwan(19), Tunisia(6), Turkey(13), UnitedKingdom(32), UnitedStates(1668), Uruguay(52) \\ \hline
  \end{tabular}%
  }
  \end{center}
  \vspace{-4mm}
  \label{tab:ips_country_akamai}%
\end{table}%

First, we observe that large countries with a relatively heavy weight in the geopolitical environment such as the US, the UK, France, Germany and Italy have a high number of Akamai nodes (i.e., associated IPs) that serve a large number of countries. The four European countries mainly serve nodes  from all over Europe, in minor level other nearby non-European countries like Israel, Jordan, Tunisia and Turkey, and on a very small scale US and China.
We also found a similar pattern in Netherlands, though it has a lower Akamai presence.
Furthermore, we discovered more Akamai nodes in US than in the rest of the countries together. These servers process queries from users located all over the world.
This will clearly have an impact on the delay for those countries that access Akamai nodes in US for a large portion of their queries, despite being far from US (e.g. Uruguay, Argentina, China and Korea).
Next, we observe that Akamai nodes in Northern European countries (Norway, Finland and Sweden) mainly respond to the demands of users located within those northern countries. A third observation is that Ireland and New Zealand should actually be located at the top of the table since they mostly attend to local FB demand, along with a few queries from China. Fourth, Akamai nodes located in small Central European countries such as Austria, Czech Republic and Switzerland, service FB demand mainly from local and nearby countries users. We can find a similar pattern for Japan and Brazil, and additionally in Australia, where the nodes mostly deal with internal demand for FB services but also receive some queries from nodes located in Japan and Taiwan. Finally, Singapore (the 4th country in terms of the number of Akamai IPs) presents the more rare results. On the one hand,  Akamai nodes in Singapore exhibit an expected behavior by serving users located in Asia. On the other hand, we discovered a very strange pattern in which Akamai nodes in Singapore attend quite a few nodes located all over America (including North and South-America).

In summary, we can conclude that FB subscribers queries are usually attended by Akamai nodes located either locally or in some nearby country. This provides a bounded access delay leading to the result presented in Section \ref{sec:access_delay} that indicates a delay that is 2.5$\times$ lower when a FB query is resolved by an Akamai node instead of a native FB server. However, we can still find some odd cases where source nodes are accessing Akamai nodes located far away which has a harmful impact in their access delay to FB services.


\section{Related Work}
We found a number of related works to our article that can be classified into two different categories: CDN Infrastructure Analysis and Facebook Services Analysis.

\textbf{CDN Infrastructure Analysis.}
There are some prior studies that analyzed different aspects of large CDNs like Akamai \cite{Akamai_Network, akamai_Drafting} or the CDN used by Google to serve youtube videos \cite{torres2011dissecting}. In the latter study the authors aim at understanding from where videos are served from, and how effective is this distribution. 
One of the main conclusions of this study is that Round-Trip Time (RTT) is used to select the preferred data center to serve the video. The studies on Akami CDN goes from a general overview \cite{Akamai_Network} to a more detailed analysis of Akamai's system components and architecture \cite{akamai_Drafting} in which authors probe Akamai network from 140 PlanetLab nodes during two months and characterize some aspects of Akamai architecture deployment such as server diversity, redirection dynamics and latency.  
Finally, we found a study \cite{FB_CDN} in which the authors examined how CDNs are used to host and serve FB content from a network perspective. This work relies on a dataset including one month of HTTP traces collected in mid 2013 from the 3G mobile network of a large European ISP. 

\textbf{Facebook Services Analysis.}
There also exist some research works that carried out different performance analyses on Facebook Services. 
Authors in \cite{wongyai2012examining} look at the established connections when FB users login in the system. In particular, they identify different sections in the FB wall page of a user, and analyze how the information filling those sections is retrieved.
An earlier work from 2010 \cite{wittie2010exploiting} identified some performance degradation (e.g., delay, packet losses, etc.) for some users accessing FB from outside US. 
Finally, we have found another interesting study \cite{beaver2010finding} that states that photo view is the most critical service for FB, and presents a detailed description on how FB photos are distributed to CDN Akamai servers. However, it does not perform a geographical analysis to understand how different regions of the world are being served as we do in our article.

\section{Lessons learned and Recommendations}

In this Section we present the most important lessons extracted from our work and provide some recommendations that could improve the currently delay experienced by users in some relevant countries.

\noindent{1. Our study confirms the good work Akamai is doing for a large-scale web service such as Facebook
. Our results show that FB is reducing the delay 2.5$\times$ by using the Akamai nodes. This latency reduction is of great importance for Facebook or any other internet service given the monetary implications associated the delay experience by end users \cite{bruce_delay}.}

\noindent{2. At the time of our study, Akamai nodes were mostly responsible of serving heavy content associated mainly to photos and videos shared in Facebook. In contrast, Facebook native servers were in charge of processes like the registration and authentication.}

\noindent{3. Akamai is very efficient ($<$50ms delay) on serving Facebook content in Europe and North-America that is explained by two factors: $(i)$ Akamai is very well positioned there with a huge number of servers. $(ii)$ A major part of the revenues obtained by FB out of advertisement happens in Europe and North-America, thus it is very important to offer a good quality of service to the subscribers in those locations.}

\noindent{4. There is some room for improving the current Facebook infrastructure in some countries like Australia and Brazil. These two mentioned countries count with 13M and 86M subscribers respectively according to the data reported by the FB ads planner, and experience a much higher access delay (87ms and 78ms respectively) than other countries with a much lower number of subscribers like Slovenia.  Therefore, we believe that Facebook should find a solution to improve the experience of Australian and Brazilian users by further exploiting Akamai nodes in those countries.}

\color{black}


\section{Conclusions}
This study presents a comprehensive measurement-based analysis of the FB network infrastructure with a special emphasis on depicting how Akamai nodes replying to FB queries from subscribers are distributed throughout the world. In this context, we have analyzed what is the average access delay that FB subscribers experience to access FB services delivered from native FB servers as well as Akamai servers depending on the country they are located. Moreover, we have thoroughly discussed the coverage offered by those Akamai nodes serving FB services.
\section{Acknowledgments}
This works is partially supported by the European Celtic-Plus project CONVINcE and eCOUSIN (EUFP7-318398).
This work is also funded by the Ministerio de Economía y Competitividad of SPAIN through the project BigDatAAM (FIS2013-47532-C3-3-P).


\end{document}